\documentclass[conference]{IEEEtran}

\usepackage{acronym}
\usepackage{amsmath,amssymb}
\usepackage{bbm}
\usepackage{caption}
\usepackage{enumerate}
\usepackage{etoolbox}
\usepackage{graphicx}
\usepackage[utf8]{inputenc}
\usepackage{lipsum}
\usepackage{siunitx}

\newtheorem{proposition}{Proposition}

\newcommand{\kr}{\ensuremath{\kappa_r}}
\newcommand{\kad}{\ensuremath{\kappa_{a_0}}}
\newcommand{\ka}{\ensuremath{\kappa_a}}
\newcommand{\kd}{\ensuremath{\kappa_d}}
\newcommand{\xmin}{\ensuremath{x_{\mathrm{min}}}}
\newcommand{\xmax}{\ensuremath{x_{\mathrm{max}}}}
\newcommand{\Cb}{\bar{C}}
\newcommand{\Ub}{\bar{U}}
\newcommand{\Wb}{\bar{W}}

\acrodef{DMC}{Diffusive Molecular Communication}
\acrodef{EH}{Energy Harvesting}
\acrodef{CIR}{channel impulse response}
\acrodef{ISI}{inter-symbol interference}
\acrodef{MC}{Molecular Communication}
\acrodef{BMI}{brain-machine interface}
\acrodef{GABA}{$\gamma$-aminobutyric acid}

\makeatletter
\long\def\@makecaption#1#2{\ifx\@captype\@IEEEtablestring%
    \footnotesize\begin{center}{\normalfont\footnotesize #1}\\
        {\normalfont\footnotesize\scshape #2}\end{center}%
    \@IEEEtablecaptionsepspace
    \else
    \@IEEEfigurecaptionsepspace
    \setbox\@tempboxa\hbox{\normalfont\footnotesize {#1.}~~ #2}%
    \ifdim \wd\@tempboxa >\hsize%
    \setbox\@tempboxa\hbox{\normalfont\footnotesize {#1.}~~ }%
    \parbox[t]{\hsize}{\normalfont\footnotesize \noindent\unhbox\@tempboxa#2}%
    \else
    \hbox to\hsize{\normalfont\footnotesize\hfil\box\@tempboxa\hfil}\fi\fi}
\makeatother

\begin{document}
\bstctlcite{disable_url}
\title{Channel Modeling for Synaptic Molecular Communication With Re-uptake and Reversible Receptor Binding}
\author{ 
	\IEEEauthorblockN{Sebastian Lotter, Arman Ahmadzadeh, and Robert Schober} 
	\IEEEauthorblockA{Friedrich-Alexander University Erlangen-Nuremberg, Germany} 
}

\maketitle
\begin{abstract}
    In \ac{DMC}, information is transmitted by diffusing molecules.
    Synaptic signaling is a natural implementation of this paradigm.
    It is responsible for relaying information from one neuron to another, but also provides support for complex functionalities, such as learning and memory.
    Many of its features are not yet understood, some are, however, known to be critical for robust, reliable neural communication.
    In particular, some synapses feature a re-uptake mechanism at the presynaptic neuron, which provides a means for removing neurotransmitters from the synaptic cleft and for recycling them for future reuse.
    In this paper, we develop a comprehensive channel model for synaptic \ac{DMC} encompassing a spatial model of the synaptic cleft, molecule re-uptake at the presynaptic neuron, and reversible binding to individual receptors at the postsynaptic neuron. 
    Based on this model, we derive an analytical time domain expression for the \ac{CIR} of the synaptic \ac{DMC} system.
    Our model explicitly incorporates macroscopic physical channel parameters and can be used to evaluate the impact of re-uptake, receptor density, and channel width on the \ac{CIR} of the synaptic \ac{DMC} system.
    Furthermore, we provide results from particle-based computer simulation, which validate the analytical model.
    The proposed comprehensive channel model for synaptic \ac{DMC} systems can be exploited for the investigation of challenging problems, like the quantification of the \acl{ISI} between successive synaptic signals and the design of synthetic neural communication systems.
\end{abstract}

\section{Introduction}
    \acresetall
    In nature, information exchange between many different biological entities, such as cells, organs, or even different individuals, is based on the release, propagation, and sensing of molecules.
    This process is called \ac{MC}.
    Although traditionally studied by biologists and medical scientists, it has recently also attracted interest in the communications research community \cite{nakano13}.
    \ac{MC} is envisioned to open several exciting new application areas for which communication at nano-scale is key \cite{akyildiz15}.
    One of these applications is the deployment and operation of synthetic cells in the human body for the purpose of tumor detection or treatment \cite{freitas99}, another one is the development of \acp{BMI} for detection or replacement of dysfunctional neural units \cite{veletic2019}.
    In both cases, communication at cell level, either between synthetic cells or between synthetic and natural cells, is required, but can not be implemented using traditional wireless communication systems.
    
    \ac{MC} systems in which molecules propagate via Brownian motion, referred to as \acf{DMC} systems, provide a viable alternative, as neither special infrastructure, nor external energy supply is needed.
    The design of synthetic \ac{DMC} systems, however, poses several challenges.
    Firstly, in environments for which synthetic \ac{DMC} is intended, such as the human body, the amount of molecules available for transmission is typically limited.
    Thus, the employed communication scheme needs to be extremely energy efficient.
    The concept of \ac{EH} enables ultra-low-power applications in traditional communications \cite{ng13} and several methods to leverage it also for \ac{DMC} have been proposed recently \cite{deng17,guo18}.
    The models considered in \cite{deng17,guo18}, however, are based on free-space propagation and thus do not take into account a particular communication environment.
    Secondly, because Brownian motion is an undirected propagation mechanism, \ac{DMC} channels are typically dispersive and adequate measures to mitigate \ac{ISI} need to be taken.
    Several approaches for \ac{ISI} mitigation in \ac{DMC} have been proposed in the literature, including ISI-aware modulation schemes \cite{arjmandi17}, forward error-correction codes \cite{leeson12,shih13}, and equalization techniques \cite{tepekule15}.
    
    A difficulty in resolving these issues is that existing concepts from traditional communications cannot be easily transferred to \ac{MC} due to limited processing capabilities at both transmitter and receiver.
    On the other hand, natural \ac{DMC} systems have evolved over millions of years to deal with such challenges.
    
    Inspired by such natural systems, enzymatic degradation of information molecules in the channel is considered for the mitigation of \ac{ISI} in \cite{noel14,heren15}.
    This approach is interesting, because it directly impacts the channel characteristics and does not increase the complexity of the transmitter or receiver.
    It does, however, incur higher energy cost for the production of enzymes and additional signaling molecules, and is thus not necessarily energy efficient.
    
    While the approach in \cite{noel14} was inspired by the neuromuscular junction, we consider a different natural \ac{DMC} system in this paper, namely molecular synaptic transmission between two neurons.
    Abstracted in communication terms, here, the presynaptic neuron (transmitter) encodes a sequence of electrical impulses (data stream) into a spatio-temporal neurotransmitter release pattern (molecular signal) which propagates through the synaptic cleft (channel) and is finally received by the postsynaptic neuron (receiver) where the neurotransmitters activate membrane receptors, see Fig.~\ref{fig:channel}.
    In addition, transporter proteins at the presynaptic neuron provide a re-uptake mechanism for many common neurotransmitters, such as dopamine, serotonin, norepinephrine, \ac{GABA}, and glycine \cite{kristensen11}.
    In this way, the channel is cleared and signaling molecules are recycled for future reuse.
    It is known that this re-uptake mechanism is critical for synaptic communication; several severe mental diseases, including attention deficit hyperactivity disorder and epilepsy are associated with dysfunctional re-uptake \cite{kristensen11}.
    
    Despite its importance for neural information transmission and its bio-physical characteristics, most existing models of synaptic signaling are based on free-space propagation \cite{balevi13,veletic15} and can thus not capture the impact of specific synaptic channel parameters.
    Recent progress in this direction has been reported in \cite{khan2017}.
    In this article, a new analytical model for molecule diffusion in the synaptic cleft, incorporating a simplified geometric representation of the synaptic cleft as infinite region bounded by to parallel planes, is proposed.
    The model in \cite{khan2017} does also take into account molecule re-uptake and binding to postsynaptic receptors, but the iterative scheme used to determine postsynaptic binding does not lend itself to analytical investigations.
    A closed-form expression for the \ac{CIR} is not provided.
    Furthermore, to evaluate the synaptic \ac{CIR} with respect to \ac{ISI}, molecule rebinding at the postsynaptic neuron, i.e., {\em reversible} binding, needs to be considered, while \cite{khan2017} assumes {\em irreversible} binding.
    
    To the best of the authors' knowledge, there is no channel model available for the synaptic cleft, which allows to assess the quantitative impact of molecule re-uptake and other channel parameters on the \ac{CIR}.
    Such a model would allow the investigation of the relevance of molecule re-uptake for \ac{EH} and \ac{ISI} mitigation, and on the other hand, be of interest in its own right, e.g.~for the development of \aclp{BMI}.
    
    The main contribution of this paper is an analytical time domain expression for the \ac{CIR} of the synaptic cleft which incorporates biologically plausible models of presynaptic re-uptake, postsynaptic reversible binding kinetics, and the particular cleft geometry.
    Furthermore, this expression is validated by particle-based computer simulation and experimentally shown to provide an unbiased estimator for the \ac{CIR}.
    
    The remainder of this paper is organized as follows:
    In Section~\ref{sec:model}, we state the system model and the main assumptions used in Section~\ref{sec:cir} to derive the \ac{CIR} of the synaptic cleft.
    The particle-based simulator design is outlined in Section~\ref{sec:pb_sim}, and numerical results are presented in Section~\ref{sec:results}.
    Finally, the main findings are summarized in Section~\ref{sec:summary}.

\section{System Model}
\label{sec:model}
\subsection{Geometry and Assumptions}
\begin{figure*}
        \centering
        \includegraphics[width=.65\textwidth]{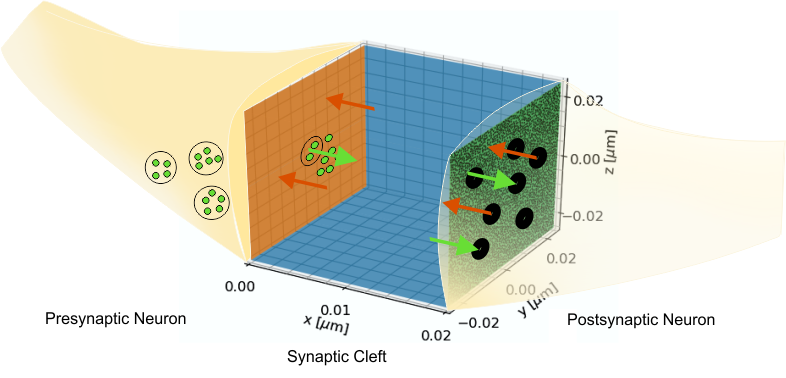}\vspace*{-3 mm}
        \caption{Model synapse. Neurotransmitters enclosed in vesicles at the presynaptic neuron are released into the synaptic cleft, propagate by Brownian motion, and activate receptors at the postsynaptic neuron. Binding to postsynaptic receptors is reversible. Furthermore, particles are re-uptaken (and recycled) at the presynaptic neuron. The cuboid represents an abstraction from the synaptic cleft. Its orange and green surfaces represent the membrane of the pre- and postsynaptic neuron, respectively.}\vspace*{-6 mm}
        \label{fig:channel}
    \end{figure*}

    The complex geometry of the synaptic cleft is abstracted in our model as a 3-dimensional rectangular cuboid, with faces in $x$-direction representing the membranes of the pre- and postsynaptic neurons (see Fig.~\ref{fig:channel}).
    This model is similar to the geometric model used in \cite{khan2017}, with the difference that the model considered here is bounded in all dimensions, while in \cite{khan2017}, $y$ and $z$ extend to infinity.
    This abstraction is analytically more tractable than the actual non-regular shaped synaptic domain, while still retaining its characteristic features.
    In the context of this model, we will refer to the boundary representing the pre- and postsynaptic membranes as the {\em left} and {\em right} boundaries, respectively.
    Formally, we denote the domain of the synaptic cleft in Cartesian coordinates as 
    \begin{multline}
    \Omega = \{(x,y,z) \vert x_{\mathrm{min}} \leq x \leq x_{\mathrm{max}},\\ y_{\mathrm{min}} \leq y \leq y_{\mathrm{max}}, z_{\mathrm{min}} \leq z \leq z_{\mathrm{max}} \}, \label{eq:domain}
    \end{multline}
    and the concentration of molecules in $\si{\per\cubic\micro\meter}$ at any time $t$ at any location within the box defined by \eqref{eq:domain} as $C_{\Omega}(x,y,z,t)$.
    
    To derive the (dimensionless) impulse response of the synaptic channel, $h(t)$, we consider instantaneous release of $N$ particles at time $t=t_0$ at location $(x,y,z)=(x_0,y_0,z_0)\in \Omega$.
    Without loss of generality (w.l.o.g.), we set $t_0=0$.
    $h(t)$ is then given as the number of particles adsorbed to the receptors of the postsynaptic membrane as a function of $t,t \geq 0$.
    For our analysis, we make four assumptions:
    \begin{enumerate}[{A}1)]
        \item The faces in $y$ and $z$ direction are fully reflective. \label{ass:1}
        \item The receptors at the postsynaptic membrane are uniformly distributed. \label{ass:2}
        \item The receptors cannot be occupied, i.e., multiple particles may bind to a single receptor. \label{ass:3}
        \item Reversible adsorption to individual receptors with intrinsic association coefficient $\kad$ in $\si{\micro\meter\per\micro\second}$ and intrinsic dissociation rate $\kd$ in $\si{\per\micro\second}$ can be treated equivalently as reversible adsorption to a homogeneous surface with effective association coefficient $\ka$ in $\si{\micro\meter\per\micro\second}$ and dissociation rate $\kd$. \label{ass:4}
    \end{enumerate}
    A\ref{ass:1} can be justified for closed neural environments, in which neurotransmitters cannot leave the synapse (no spill-over to other synapses).
    A\ref{ass:2} is reasonable as long as only the so-called postsynaptic density \cite{sheng07} is considered, the part of the postsynaptic membrane which contains most receptors.
    A\ref{ass:3} can be justified for low molecule concentrations.
    Finally, A\ref{ass:4} is not obvious.
    For irreversible adsorption to a otherwise reflective surface covered by partially adsorbing disks, boundary homogenization has been justified in \cite{zwanzig1991} and refined with results from computer simulation in \cite{berezhkovskii04}.
    However, it is not clear if similar techniques can be applied for reversible reactions, too.
    There are two main issues to consider here.
    First, the steady-state fluxes are substantially different;
    the net flux at a reversibly adsorbing boundary is $0$ at steady-state, while for irreversible adsorption it is nonzero.
    Second, desorption alters the spatial concentration profile of particles near the boundary; particles are more concentrated near receptors compared to irreversible adsorption \cite{szabo91}.
    The first issue can be resolved by calibrating the effective adsorption coefficient to the homogenized surface such that it correctly reproduces the steady-state flux to the patchy surface (see Section~\ref{sec:pb_sim}), while the second issue can be resolved in a biologically plausible manner by assuming that particles may not re-adsorb immediately after unbinding.
    
    With these assumptions, $h(t)$ becomes independent of the particle distribution in $y$ and $z$, and, instead of $C_{\Omega}(x,y,z,t)$, it is sufficient to consider the concentration of molecules aggregated over $y$ and $z$, $C(x,t)$, in $\si{\per\micro\meter}$, where
    \begin{equation}
    C(x,t) =  \int\limits_{y_{\mathrm{min}}}^{y_{\mathrm{max}}}\int\limits_{z_{\mathrm{min}}}^{z_{\mathrm{max}}} C_{\Omega}(x,\nu,\xi,t) \mathrm{d}\xi \mathrm{d}\nu.
    \end{equation}
    
    According to Fick's second law of diffusion, (average) Brownian particle motion can be described by the following partial differential equation:
    \begin{equation}
    \frac{\partial{}C}{\partial{}t} = D\frac{\partial{}^2C}{\partial{}x^2}, \label{eq:diff_oned}
    \end{equation}
    where $D$ denotes the particle diffusion coefficient in $\si{\micro\meter\squared\per\micro\second}$.
    
\subsection{Pre- and Postsynaptic Neurons}
    Particle re-uptake at the presynaptic neuron can be modeled as irreversible adsorption to the homogeneous left boundary at $x=\xmin$ (possibly after appropriate boundary homogenization) with re-uptake coefficient $\kr$ in $\si{\micro\meter\per\micro\second}$.
    According to the classical result in \cite{collins1949}, adsorption of particles to a partially adsorbing boundary can be modeled with a radiating boundary condition.
    To simplify notation, we set $\xmin=0$ and $\xmax=a$.
    Then, the radiating boundary condition modeling presynaptic re-uptake is given as
    \begin{equation}
    D\frac{\partial C}{\partial x} =  \kr C, \text{ at } x=0. \label{eq:pab_lb}
    \end{equation}
    
    For the right boundary, the radiating boundary needs to be extended to incorporate particle desorption.
    We follow a similar approach as described in \cite{ahmadzadeh16}.
    Namely, particle desorption is modeled as a first-order process depending on the intrinsic desorption rate $\kd$ and on the amount of currently adsorbed particles, which, in turn, corresponds to $h(t)$.
    Thus, the boundary condition for reversible adsorption at the right boundary can be formulated as
    \begin{align}
    D\frac{\partial C}{\partial x} &= -\overbrace{\ka C}^{\text{adsorption}} + \overbrace{\kd h(t)}^{\text{desorption}}, \textrm{ at } x = a\\
    \frac{\mathrm{d} h(t)}{\mathrm{d} t} &= \ka C(a,t) - \kd h(t),
    \end{align}
    which simplifies directly to
    \begin{equation}
    D\frac{\partial C}{\partial x} = - \kappa_a C - \kappa_d \int_{0}^{t} D\frac{\partial C}{\partial x} \mathrm{d}\tau, \text{ at } x = a. \label{eq:pab_rb}
    \end{equation}
    
    Finally, instantaneous release of $N$ particles at $t_0=0$ is modeled with the initial value
    \begin{equation}
    C(x,0) = N \delta(x-x_0), 0 \leq x_0 \leq a, \label{eq:pab_ic}
    \end{equation}
    where $\delta(x)$ denotes the Dirac delta function.
    For the following derivations, we set w.l.o.g.~$N=1$.
    
\subsection{Channel Impulse Response}
    Once the solution to \eqref{eq:diff_oned}, \eqref{eq:pab_lb}, \eqref{eq:pab_rb}, \eqref{eq:pab_ic} is found, $h(t)$ can be obtained as
    \begin{equation}
    h(t) = \int\limits_{0}^{t} -D\left.\frac{\partial C(x,\tau)}{\partial x}\right|_{x=a} \mathrm{d}\tau. \label{eq:pab:ht_def}
    \end{equation}
    
\section{Analytical Channel Model}
\label{sec:cir}

\subsection{Molecule Concentration}
    To find the $C(x,t)$ that satisfies \eqref{eq:diff_oned}, \eqref{eq:pab_lb}, \eqref{eq:pab_rb}, \eqref{eq:pab_ic}, we use a similar approach as \cite[Ch.~14]{carslaw86} and decompose $C$ as 
    \begin{equation}
    C = U + W,\label{eq:C_dec}
    \end{equation}
    such that $U$ fulfills \eqref{eq:diff_oned} and \eqref{eq:pab_ic}, $W$ fulfills \eqref{eq:diff_oned} and equals $0$ at $t_0$, and $U$ and $W$ together fulfill \eqref{eq:pab_lb} and \eqref{eq:pab_rb}.
    Next, we assume that the Laplace transform of $C(x,t)$ with respect to (w.r.t.)~$t$, $\Cb(x,p)=\mathcal{L}\{C(x,t)\}=\int_{0}^{\infty}C(x,\tau)\exp(-p\tau) \mathrm{d}\tau$, exists.
    With this assumption and \eqref{eq:C_dec}, $C(x,t)$ can be found.

    \begin{proposition}        
        Let $\kr,\kd \geq 0$, $\ka > 0$\footnote{For $\ka = 0$, $h(t) = 0$, therefore this case is not considered here.}.
        The unique solution to \eqref{eq:diff_oned}, \eqref{eq:pab_lb}, \eqref{eq:pab_rb}, \eqref{eq:pab_ic} is
        \begin{equation}
        C(x,t) = \sum_{n = 1}^{\infty} Z_n(x)Z_n(x_0)e^{-D \alpha_n^2 t} + \frac{\kappa_d}{\kappa_a + a \kappa_d}\mathbbm{1}(\kr = 0), \label{eq:pab_sol_sum}
        \end{equation}
        where $\mathbbm{1}(\cdot)$ denotes the indicator function,
        \begin{multline}
        Z_n(x) = \frac{[2 \left(\alpha _n^2 \left(\kappa _a^2-2 D \kappa _d\right)+\kappa _d^2+D^2 \alpha _n^4\right)]^{\frac{1}{2}} }{\mathcal{D}^{\frac{1}{2}}}\\
        \times \left(D \alpha _n \cos \left(x \alpha _n\right)+\kappa _r \sin \left(x \alpha _n\right)\right),
        \end{multline}
        \begin{multline}
        \mathcal{D} = a \left(D^2 \alpha _n^2+\kappa _r^2\right) \left(\alpha _n^2 \left(\kappa _a^2-2 D \kappa _d\right)+\kappa _d^2+D^2 \alpha _n^4\right)\\
        +D \alpha _n^2 \left(D \kappa _d \left(\kappa _a-2 \kappa _r\right)+\kappa _a \kappa _r \left(\kappa _a+\kappa _r\right)\right) \\
        +\kappa _d \kappa _r \left(\kappa _a \kappa _r+D \kappa _d\right)+D^3 \alpha _n^4 \left(\kappa _a+\kappa _r\right), \label{eq:def_D}
        \end{multline}
        
        and the $\alpha_n$ are defined as the positive roots of
        \begin{equation}
        \tan (a \alpha) = \frac{D \left(\kappa_a + \kappa_r \right) \alpha^2 - \kappa_d \kappa_r}{D^2 \alpha^3 - \left(\kappa_a \kappa_r + D \kappa_d\right)\alpha}.
        \end{equation}
    \end{proposition}

    \begin{IEEEproof}
    Please refer to the Appendix.
    \end{IEEEproof}

    For {\em irreversible} adsorption to the right boundary (i.e., $\kappa_d = 0$), \eqref{eq:pab_sol_sum} reduces to \cite[Ch.~14.3, eq.~(4)]{carslaw86}, from which, in turn, follows \cite[eq.~(5)]{khan2017} after multiplying with the Green's function for free diffusion in $y$ and $z$.
    From now on, we assume $\kappa_d > 0$.
    
    If there is no re-uptake, i.e.,~$\kappa_r = 0$, the steady-state concentration for large $t$ is 
    \begin{equation}
    \frac{\kappa_d}{\kappa_a + a \kappa_d} \label{eq:pab_res_h1_0},
    \end{equation}
    because all exponential terms in \eqref{eq:pab_sol_sum} vanish.
    This is intuitive, as in the absence of re-uptake, particles are not ultimately removed from the system.
    Also, the steady-state is a well-mixed state with constant concentration everywhere, independent of $x$.
    The net flux at the right boundary in this case is zero, meaning at each time step, the same number of particles adsorb and desorb.
    Integrating over the size of the cleft, $a$, the fraction of solute particles in the steady-state is then given as
    \begin{equation}
    \frac{a \kappa_d}{\kappa_a + a \kappa_d}, \label{eq:pab_c_steadystate_h1_0}
    \end{equation}
    while the fraction of permanently adsorbed particles is given as
    \begin{equation}
    1 - \frac{a \kappa_d}{\kappa_a + a \kappa_d} = \frac{\kappa_a}{\kappa_a + a \kappa_d}.\label{eq:c_inf_h1_0_adsorbed}
    \end{equation}
    
    This result is a special case of \cite[eq.~(2.5)]{berezhkovskii13a}.
    If $a\kappa_d \gg \kappa_a$, \eqref{eq:pab_c_steadystate_h1_0} approaches $1$, i.e., almost all particles are solute, while, if $\kappa_a \gg a\kappa_d$, the concentration of solute particles approaches $0$ and almost all particles are bound to receptors in the steady-state.
    
    In the presence of re-uptake (i.e.~$\kappa_r > 0$), in contrast, there is no constant term in \eqref{eq:pab:ht}, meaning that the concentration of particles for $t \to \infty$ approaches $0$ everywhere.
    Again, this is intuitive, as in the presence of re-uptake, all particles are eventually re-uptaken.
    
\subsection{Channel Impulse Response}
    To obtain $h(t)$, we differentiate \eqref{eq:pab_sol_sum} with respect to $x$, and reach
    \begin{equation}
    -\frac{\partial C(x,t)}{\partial x} = \sum_{n = 1}^{\infty} Z'_n(x)Z_n(x_0)e^{-D \alpha_n^2 t},
    \end{equation}
    where
    \begin{multline}
    Z'_n(x) = \frac{[2 \left(\alpha _n^2 \left(\kappa _a^2-2 D \kappa _d\right)+\kappa _d^2+D^2 \alpha _n^4\right)]^{\frac{1}{2}} }{\mathcal{D}^{\frac{1}{2}}}\\
    \times \left(D \alpha_n^2 \sin \left(x \alpha _n\right)-\kappa_r \alpha_n \cos \left(x \alpha _n\right)\right). \label{eq:pab:diff_Z_x}
    \end{multline}
    
    Evaluating \eqref{eq:pab:diff_Z_x} at $x=a$, assuming we may interchange integration and summation in \eqref{eq:pab:ht_def}, and computing the integral yields for the \ac{CIR}
    \begin{equation}
    h(t) = \sum_{n = 1}^{\infty} Z'_n(a)Z_n(x_0) \frac{\left(1-e^{-D \alpha_n^2 t}\right)}{\alpha_n^2}. \label{eq:pab:ht}
    \end{equation}
    
    Due to the differentiation in \eqref{eq:pab:ht_def}, this term is independent of the constant term in \eqref{eq:pab:ht} and thus valid for all $\kr \geq 0$.
    
    We note that the sequence $(\alpha_n)_n$ is strictly monotonically increasing.
    Therefore, for large $t$, the contributions of large-$n$ terms are (almost) constant.
    In fact, the tail of $h(t)$ can be properly approximated with
    \begin{equation}
    h_{1}(t) = -Z'_1(a)Z_1(x_0) \frac{e^{-D \alpha_1^2 t}}{\alpha_1^2}, \label{eq:pab:ht_fta}
    \end{equation}
    see also Figs.~\ref{fig:coverage_increase}--\ref{fig:cleft_width_increase}.
    Such an approximation can be useful for detector design and also a first step in quantifying \ac{ISI}.
    However, due to space constraints, further analytical investigations are left for future work.

\section{Particle-based Simulation}
\label{sec:pb_sim}
    To verify the analytical expression derived in Section~\ref{sec:cir}, 3-dimensional particle-based computer simulations were conducted.
    To this end, we adopted the simulator design from \cite{andrews09}.
\subsection{Simulator Design}
    Here, Brownian particle motion is simulated by updating the position of each particle at each time step with a 3-dimensional jointly independent Gaussian random vector
    \begin{equation}
    [X, Y, Z] \sim \mathcal{N}(\mathbf{0}_{1 \times 3},\sigma^2 \mathbf{I}_{3 \times 3}),
    \end{equation}
    where $\mathcal{N}(\boldsymbol\mu,\boldsymbol{\mathcal{C}})$ denotes a multivariate Gaussian distribution with mean vector $\boldsymbol\mu$ and covariance matrix $\boldsymbol{\mathcal{C}}$, and $\mathbf{0}_{1 \times M}$ and $\mathbf{I}_{M \times M}$ denote the $1 \times M$ all-zero vector and the $M \times M$ identity matrix, respectively.
    Particle collisions are neglected.
    The variance $\sigma^2$ is a function of the particle diffusion coefficient $D$ and the simulation time step $\Delta t$,
    \begin{equation}
    \sigma^2 = 2 D \Delta t.
    \end{equation}
    
    Accordingly, the {\em root mean step length (rms)} of a simulated particle is defined as
    \begin{equation}
    s = \sqrt{2 D \Delta t}.
    \end{equation}
    
    For the simulation, the probability that a particle is re-uptaken after crossing the left boundary was computed from $\kr$ using \cite[eq.~(21)]{andrews09}.
    Particles hitting a receptor at the postsynaptic boundary were absorbed with probabilities computed as in \cite[eqs.~(37), (32)]{andrews09} from the intrinsic association coefficient of molecules to receptors, $\kad$, and the intrinsic desorption rate constant, $\kd$.

    \subsection{Boundary Homogenization}

    In order to compare simulation data and analytical results, boundary homogenization at the right boundary needs to be performed.
    To avoid issues that arise from the non-uniform distribution of desorbed particles, we choose $\Delta t$ such that $s$ is larger than the receptor radius, $r$.
    For fixed surface coverage at the postsynaptic neuron, increasing $\Delta t$ has the effect that desorbing particles are more likely to see a representative part of the boundary before possible adsorbing to a receptor again.
    This is in fact equivalent to blocking the desorbed particle for some time from re-adsorbing and thus fulfills one of the conditions of A\ref{ass:4}.
    Now, to compute the effective adsorption coefficient for the homogenized boundary, $\ka$, from $\kad$, we perform computer-assisted boundary homogenization similar to \cite{berezhkovskii04}, with the difference, that we do not require analytical and numerical results to produce the same {\em average particle life times}, but instead demand matching {\em steady-state concentrations}.
    To this end, we conduct particle-based simulations without re-uptake and let them run into steady-state.
    Next, we use our analytical result for the steady-state number of adsorbed particles in the absence of re-uptake, \eqref{eq:c_inf_h1_0_adsorbed}, to fit $\ka$.
    In contrast to \cite{berezhkovskii04} and earlier results, it turns out that $\ka$ for reversible reaction varies only moderately with $r$, if $r$ is close to $s$, and mostly depends on the fraction of the postsynaptic surface covered by receptors, $\rho$, and the intrinsic receptor association coefficient $\kad$.
    We found that
    \begin{equation}
        \ka = 0.984 \rho \kad
    \end{equation}
    provides a good approximation for receptors of radius $r= \SI{0.3}{\nano\meter}$, which is slightly less than $s$ for $\Delta t = \SI{1}{\nano\second}$ and $D = \SI{6.8e-5}{\micro\meter\squared\per\micro\second}$.

\section{Simulation Results}
\label{sec:results}
    To simplify the interpretation of the simulation results presented in this section, some of the simulation parameters are given as dimensionless quantities \cite{andrews09}.
    To this end, we define the {\em reduced re-uptake coefficient $\kr'$} \cite[eq.~(12)]{andrews09} as
    \begin{equation}
        \kr' = \frac{\kr\sqrt{\Delta t}}{\sqrt{2D}},
    \end{equation}
    the {\em reduced intrinsic adsorption coefficient $\kad'$} \cite[eq.~(12)]{andrews09} as
    \begin{equation}
        \kad' = \frac{\kad\sqrt{\Delta t}}{\sqrt{2D}},
    \end{equation}
    and, finally, the {\em reduced desorption rate $\kd'$} \cite[eq.~(13)]{andrews09} as
    \begin{equation}
        \kd' = \kd \Delta t.
    \end{equation}
    
    For the diffusion coefficient and the width of the synaptic cleft, we used values from the literature.
    The other parameters were varied to ensure the validity of our model for a wide range of sensible parameter values.
    A complete listing of the default parameters can be found in Table~\ref{tab:sim_params}.
    
    \begin{table}
        \vspace*{0.05in}
        \centering
        \caption{Simulation parameters for particle-based simulation \cite{rice85,alberts14}.}
        \footnotesize
        \begin{tabular}{| l | r | p{.4\linewidth} |}
            \hline Parameter & Default Value & Description \\ \hline
            $D$ & $\SI{6.8e-5}{\micro\meter\squared\per\micro\second}$& Particle diffusion coefficient \\ \hline
            $\Delta t$ & $\SI{1}{\nano\second}$ & Simulation time step \\ \hline
            $N$ & $\SI{2000}{}$ & Number of released particles \\ \hline
            $a$ & $\SI{20}{\nano\meter}$& Channel width in $x$ direction \\ \hline
            $y_{\textrm{max}}-y_{\textrm{min}}$ & $\SI{0.05}{\micro\meter}$ & Channel width in $y$ direction \\ \hline
            $z_{\textrm{max}}-z_{\textrm{min}}$ & $\SI{0.05}{\micro\meter}$ & Channel width in $z$ direction \\ \hline
            $\kappa'_r$ & $\SI{0.02}{}$ & Reduced re-uptake coefficient \\ \hline
            $\kad'$ &$\SI{1.0}{}$ & Reduced intrinsic adsorption coefficient \\ \hline
            $\kappa'_d$ &$\SI{0.7}{}$ & Reduced desorption rate \\ \hline
            $r$ & $\SI{0.3}{\nano\meter}$ & Receptor radius \\ \hline
            $\rho$ & $\SI{0.4}{}$ & Receptor coverage at postsynaptic neuron \\ \hline
        \end{tabular}\vspace*{-4 mm}
        \label{tab:sim_params}
    \end{table}

    As particles diffuse independently, the \ac{CIR} for $N > 1$ is obtained by multiplying \eqref{eq:pab:ht} by $N$.
    For the numerical evaluation of \eqref{eq:pab:ht}, the sum was truncated after the first $500$ terms.
    After proper calibration of the homogenized adsorption coefficient, $\ka$, as described in Section~\ref{sec:pb_sim}, the agreement between analytical solution and simulation results was in general excellent for all parameter regimes that were tested.
    
\begin{figure}[!t]
        \centering
        \includegraphics[width=.45\textwidth]{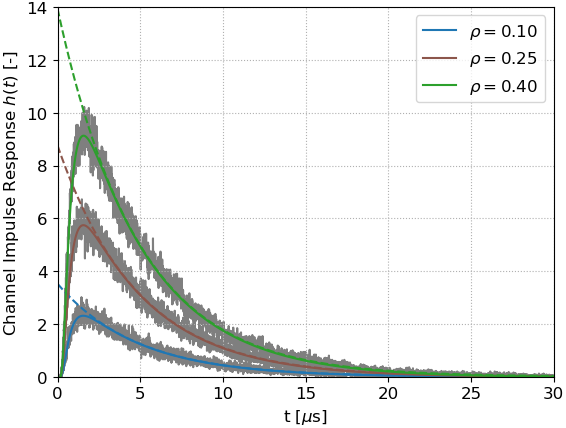}
        \caption{Results for particle-based simulation (gray curves) and analytical solutions (solid lines) for different receptor densities at the postsynaptic neuron. The first-term approximation \eqref{eq:pab:ht_fta} is shown with a dashed line.}\vspace*{-6 mm}
        \label{fig:coverage_increase}
    \end{figure}

    First, we investigate the impact of the receptor coverage at the postsynaptic neuron, $\rho$, on the \ac{CIR}.
    In Fig.~\ref{fig:coverage_increase}, it can be seen that, on the one hand, increasing the receptor coverage increases the peak value, but, on the other hand, the \ac{CIR} also takes more time to decay.
    This is expected as by increasing the receptor coverage, firstly, more particles adsorb to the postsynaptic membrane at their first arrival, and, secondly, desorbed particles are more likely to rebind immediately after desorption.
    Also, Fig.~\ref{fig:coverage_increase} shows that the \ac{CIR} can be approximated with high accuracy after its peak using the first term of the sum in \eqref{eq:pab:ht}.
    
\begin{figure}[!t]
        \centering
        \includegraphics[width=.45\textwidth]{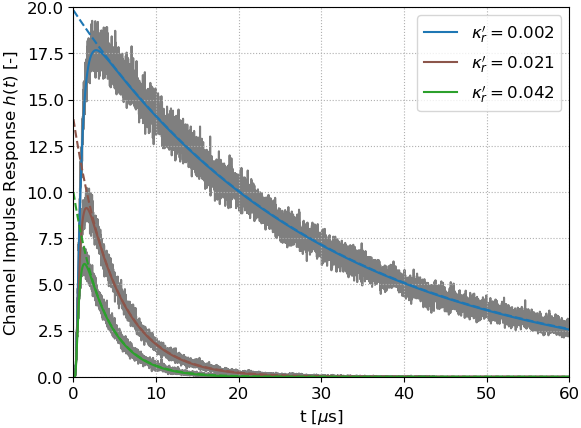}
        \caption{Results for particle-based simulation (gray curves) and analytical solutions (solid lines) are shown for different re-uptake rates $\kr$ together with the first-term approximation  \eqref{eq:pab:ht_fta} (dashed lines).}\vspace*{-3 mm}
        \label{fig:reuptake_increase}
    \end{figure}

    Next, we investigate the effect of particle re-uptake on the \ac{CIR}.
    From Fig.~\ref{fig:reuptake_increase}, it can be observed that increasing the re-uptake rate considerably shortens the \ac{CIR}.
    At the same time, however, the peak value of the received signal is reduced.
    Again, this trade-off is expected as, for larger $\kr'$, particles are more likely to be re-uptaken before they get the chance to even reach the postsynaptic side.
    
\begin{figure}[!t]
        \centering
        \includegraphics[width=.45\textwidth]{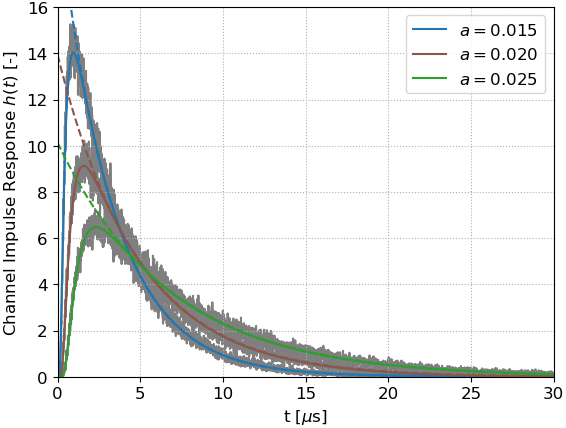}
        \caption{Results for particle-based simulation (gray curves) and analytical solutions (solid lines) are shown for different cleft widths together with the first-term approximation \eqref{eq:pab:ht_fta} (dashed lines).}\vspace*{-6 mm}
        \label{fig:cleft_width_increase}
    \end{figure}

    Finally, the impact of the channel width on the \ac{CIR} is investigated in Fig.~\ref{fig:cleft_width_increase}.
    Increasing the channel width leads to a more dispersive \ac{CIR} and a more severe attenuation of the signal.
    Interestingly, in contrast to the other parameters we investigated in Figs.~\ref{fig:coverage_increase} and \ref{fig:reuptake_increase}, here, decreasing the cleft width leads to a shorter {\em and} stronger signal.
    Thus, in terms of \ac{ISI} mitigation {\em and} peak value, a small cleft width is beneficial.

\section{Conclusions}
\label{sec:summary}

In this paper, a new analytical model for the synaptic channel has been proposed.
An analytical time domain solution for the \ac{CIR} has been derived and validated with particle-based simulation.
The supposed effect of presynaptic molecule re-uptake, namely a significant shortening of the \ac{CIR}, has been observed in the model system.
Furthermore, it has been demonstrated that molecule re-uptake also leads to a lower peak value.
It was shown that higher receptor density has a positive effect on the peak value, while, at the same time, also the undesired signal part (tail) is enhanced.
In addition, the impact of the width of the synaptic cleft on the \ac{CIR} was investigated and it was shown that a smaller distance between transmitter and receiver is beneficial.
Finally, the first term of the infinite sum in the proposed \ac{CIR} expression was shown to provide an easy-to-compute approximation for the tail of the \ac{CIR}.

We believe that by incorporating many important bio-physical features of the synaptic communication channel, the proposed model is useful for the design of future synthetic neural communication systems.

\appendix

\section{}
\subsection{Sketch of Derivation of CIR}
\label{sec:app:time_domain_derivation}

Eq.~\eqref{eq:pab_lb} reads in the Laplace domain
\begin{equation}
D\frac{\partial \bar{C}}{\partial x} = \kappa_r \bar{C}, \text{ for } x = 0, \label{eq:pab_lb_lapl}
\end{equation}
while \eqref{eq:pab_rb} transforms to
\begin{equation}
D \frac{\partial \bar{C}}{\partial x} = -\kappa_a \bar{C} - \kappa_d D \frac{1}{p} \frac{\partial \bar{C}}{\partial x}, \text{ for } x = a. \label{eq:pab_rb_lapl}
\end{equation}

Here, $p$ is the Laplace variable and we define $q = \sqrt{\frac{p}{D}}$.

Let $\bar{U}$ and $\bar{W}$ denote the Laplace transforms of $U$ and $W$, as given in \eqref{eq:C_dec}, respectively.

Following \cite[Ch.~14.3]{carslaw86}, $\bar{U}$ and $\bar{W}$ can be found to be 
\begin{equation}
\bar{U} = \frac{1}{2 D q}e^{-q|x-x'|},
\end{equation}
and
\begin{equation}
\bar{W} = A \sinh(qx) + B \cosh(qx),
\end{equation}
where constants $A$ and $B$ are to be chosen such that \eqref{eq:pab_lb_lapl} and \eqref{eq:pab_rb_lapl} are fulfilled.

Substituting $\Cb = \Ub + \Wb$ in \eqref{eq:pab_lb_lapl} and \eqref{eq:pab_rb_lapl} and solving for $A$ and $B$, we obtain $\bar{C}$ as
\begin{multline}
\bar{C}(x,p) = e^{-q \left(a+\left| x-x_0\right| \right)} \\
\left(D q e^{q \left(a+\left| x-x_0\right| -x_0\right)} \left(\cosh (q (a-x)) \left(\kappa _d+D q^2\right)\right.\right.\\ 
\left.+q \kappa _a \sinh (q (a-x))\right)
 -\kappa _d \kappa _r \cosh (q (a-x)) e^{q \left(a+\left| x-x_0\right| -x_0\right)} \\
- D q^2 \kappa _r \cosh (q (a-x)) e^{q \left(a+\left| x-x_0\right| -x_0\right)}\\
-D q^2 \kappa _a \cosh (q x) e^{q \left(\left| x-x_0\right| +x_0\right)} \\
- q \kappa _a \kappa _r \sinh (q (a-x)) e^{q \left(a+\left| x-x_0\right| -x_0\right)}\\
-q \kappa _a \kappa _r \sinh (q x) e^{q \left(\left| x-x_0\right| +x_0\right)} + D q e^{a q} \kappa _d \sinh (a q) \\
+e^{a q} \kappa _d \kappa _r \cosh (a q)+D^2 q^3 e^{a q} \sinh (a q)\\
+D q^2 \kappa _a e^{a q} \cosh (a q)+D q^2 e^{a q} \kappa _r \cosh (a q) \\
+  q \kappa _a e^{a q} \kappa _r \sinh (a q)\\
+\left.e^{q \left(\left| x-x_0\right| +x_0\right)} \left(\kappa _d+D q^2\right) \left(D q \cosh (q x)+\kappa _r \sinh (q x)\right) \right) / \\
\left[ 2 D q \left(q \sinh (a q) \left(\kappa _a \kappa _r+D \left(\kappa _d+D q^2\right)\right)\right.\right.\\
\left.\left.+\cosh (a q) \left(D q^2 \kappa _a+\kappa _r \left(\kappa _d+D q^2\right)\right)\right) \right]. \label{eq:pab_sol_lapl}
\end{multline}

\subsection{Time Domain Solution}

The corresponding solution in the time domain is now given by the inverse Laplace transform:
\begin{equation}
C(x,t) = \frac{1}{2 \pi j} \int_{\gamma -j \infty}^{\gamma + j \infty} e^{pt}\bar{C}(x,p) dp, \label{eq:inv_lapl_trans}
\end{equation}
where $j$ denotes the imaginary unit and $\gamma \in \mathbb{R}^+$ needs to be chosen large enough such that all singularities of $\bar{C}$ are left of the line along which the integral is computed.

Similarly to \cite{carslaw86}, instead of integrating along the infinite line, we complete it with a semicircle to a closed contour which contains the origin and all singularities of $\bar{C}$.
Then, we use the residue theorem to replace this contour integral with a sum over the residues $\{\sigma\}$ of $\bar{C}$,
\begin{equation}
C(x,t) = \sum_{\{\sigma\}} \mathrm{Res}(e^{pt}\bar{C},\sigma). \label{eq:residue_sum}
\end{equation}

First, we note that $\bar{C}$ can be written as quotient of two functions, $f$ and $g$, which are given as the numerator and denominator of \eqref{eq:pab_sol_lapl}, respectively.
The residues of $e^{pt}\bar{C}$ at all simple poles $\sigma$ can then be computed by l'H\^opital's rule as
\begin{equation}
\textrm{Res}\left(e^{pt}\bar{C},\sigma\right) = \frac{e^{\sigma t}f(x,\sigma)}{g'(x,\sigma)}. \label{eq:res_lhopital}
\end{equation}
For higher order poles, the residues can be found by expanding $e^{pt}\bar{C}$ as Laurent series.

Let us now look at the denominator of $\bar{C}$,
\begin{multline}
2 D q \left(q \sinh (a q) \left(\kappa_a \kappa_r+D \left(\kappa_d+D q^2\right)\right)\right.\\
\left.+\cosh (a q) \left(D q^2 \kappa_a+\kappa_r \left(\kappa_d+D q^2\right)\right)\right). \label{eq:app:den_lapl}
\end{multline}
The parameters $D$ and $a$ are always positive.

Now, let us first assume that $\kappa_r, \kappa_a$, and $\kappa_d$ are also positive. In this case, \eqref{eq:app:den_lapl} has one simple pole at $q=0$ and nonzero simple poles at the roots of
\begin{equation}
\frac{\sinh (a q)}{\cosh (a q)} = \frac{D q^2 \kappa_a+\kappa_r \left(\kappa_d+D q^2\right)}{q \left(\kappa_a \kappa_r+D \left(\kappa_d+D q^2\right)\right)}. \label{eq:app:roots_p}
\end{equation}
Set $q=j\alpha$, then the non-zero singularities of $\bar{C}$ are given as the roots $\pm\alpha_n, n \in \mathbb{N}$, of
\begin{equation}
\tan (a \alpha) = \frac{D \left(\kappa_a + \kappa_r \right) \alpha^2 - \kappa_d \kappa_r}{D^2 \alpha^3 - \left(\kappa_a \kappa_r + D \kappa_d\right)\alpha}. \label{eq:app:roots_an}
\end{equation}
$\textrm{Res}\left(e^{pt}\bar{C},0\right)$ can be computed to be
\begin{equation}
\textrm{Res}\left(e^{pt}\bar{C},0\right) = \frac{0}{2 D \kappa_r \kappa_d} = 0.
\end{equation}
By repeatedly exploiting \eqref{eq:app:roots_an} and finally plugging in $p=-D \alpha_n^2$, the residues of $e^{pt}\bar{C}$ at the simple non-zero poles $\alpha_n$ can be computed to be
\begin{multline}
\textrm{Res}\left(e^{pt}\bar{C},\alpha_n\right) = \frac{2 \left(\alpha _n^2 \left(\kappa _a^2-2 D \kappa _d\right)+\kappa _d^2+D^2 \alpha _n^4\right)}{\mathcal{D}} \\
\times \left(D \alpha _n \cos \left(x \alpha _n\right)+\kappa _r \sin \left(x \alpha _n\right)\right)\\
\times \left(D \alpha _n \cos \left(\alpha _n x'\right)+\kappa _r \sin \left(\alpha _n x'\right)\right), \label{eq:res_simp}
\end{multline}
where $\mathcal{D}$ is defined in \eqref{eq:def_D}.
Factorizing \eqref{eq:res_simp} and summing over all $\alpha_n$ yields \eqref{eq:pab_sol_sum}.

Now, if $\kappa_a > 0$ and $\kappa_d > 0$, but $\kappa_r = 0$, \eqref{eq:app:den_lapl} has a double root at $q=0$ and $\textrm{Res}\left(e^{pt}\bar{C},0\right)$ is most easily computed from the Laurent series expansion of $e^{pt}\bar{C}$.
Namely, the residue of $e^{pt}\bar{C}$ coincides with the coefficient $a_{-1}$ of its Laurent series expansion at $p=0$.

Expanding \eqref{eq:app:den_lapl} in $q$ yields
\begin{equation}
g(x,0) = q^2 (2 a D \kappa_d+2 D \kappa_a)+O\left(q^3\right). \label{eq:app:res_h1_0_den}
\end{equation}
Because $p = Dq^2$, the coefficient of $p^{-1}$ in the expansion of $e^{pt}\bar{C}$ corresponds to the coefficient of $(Dq^{2})^{-1}$ and it is clear that only coefficients of constant terms from the expansion of the numerator in $q$ play a role.
Thus, expanding the terms $\kappa_d \cosh(q (a - x))$ and $\kappa_d \cosh(q x)$ and dividing by \eqref{eq:app:res_h1_0_den} yields
\begin{equation}
\textrm{Res}\left(e^{pt}\bar{C},0\right) = \frac{\kappa_d}{\kappa_a + a \kappa_d}.
\end{equation}
For $\kd=0$, $C(x,t)$ can be obtained in a similar fashion using \eqref{eq:res_lhopital}.
This completes the proof.

\bibliographystyle{IEEEtran}
\bibliography{IEEEabrv,mc}

\begin{thebibliography}{10}
\providecommand{\url}[1]{#1}
\csname url@samestyle\endcsname
\providecommand{\newblock}{\relax}
\providecommand{\bibinfo}[2]{#2}
\providecommand{\BIBentrySTDinterwordspacing}{\spaceskip=0pt\relax}
\providecommand{\BIBentryALTinterwordstretchfactor}{4}
\providecommand{\BIBentryALTinterwordspacing}{\spaceskip=\fontdimen2\font plus
\BIBentryALTinterwordstretchfactor\fontdimen3\font minus
  \fontdimen4\font\relax}
\providecommand{\BIBforeignlanguage}[2]{{%
\expandafter\ifx\csname l@#1\endcsname\relax
\typeout{** WARNING: IEEEtran.bst: No hyphenation pattern has been}%
\typeout{** loaded for the language `#1'. Using the pattern for}%
\typeout{** the default language instead.}%
\else
\language=\csname l@#1\endcsname
\fi
#2}}
\providecommand{\BIBdecl}{\relax}
\BIBdecl

\bibitem{nakano13}
T.~Nakano, A.~W. Eckford, and T.~Haraguchi, \emph{Molecular
  Communication}.\hskip 1em plus 0.5em minus 0.4em\relax Cambridge University
  Press, 2013.

\bibitem{akyildiz15}
I.~F. {Akyildiz}, M.~{Pierobon}, S.~{Balasubramaniam}, and Y.~{Koucheryavy},
  ``The internet of bio-nano things,'' \emph{IEEE Commun. Mag.}, vol.~53,
  no.~3, pp. 32--40, Mar. 2015.

\bibitem{freitas99}
R.~A. Freitas, \emph{Nanomedicine, Volume I: Basic Capabilities}.\hskip 1em
  plus 0.5em minus 0.4em\relax Landes Bioscience Georgetown, TX, 1999, vol.~1.

\bibitem{veletic2019}
M.~{Veleti{\'{c}}} and I.~{Balasingham}, ``Synaptic communication engineering
  for future cognitive brain--machine interfaces,'' \emph{Proc. IEEE}, vol.
  107, no.~7, pp. 1425--1441, Jul. 2019.

\bibitem{ng13}
D.~W.~K. Ng, E.~S. Lo, and R.~Schober, ``Wireless information and power
  transfer: Energy efficiency optimization in {OFDMA} systems,'' \emph{IEEE
  Trans. Wireless Commun.}, vol.~12, no.~12, pp. 6352--6370, 2013.

\bibitem{deng17}
Y.~{Deng}, W.~{Guo}, A.~{Noel}, A.~{Nallanathan}, and M.~{Elkashlan},
  ``Enabling energy efficient molecular communication via molecule energy
  transfer,'' \emph{IEEE Commun. Lett.}, vol.~21, no.~2, pp. 254--257, Feb.
  2017.

\bibitem{guo18}
W.~{Guo}, Y.~{Deng}, H.~B. {Yilmaz}, N.~{Farsad}, M.~{Elkashlan}, A.~{Eckford},
  A.~{Nallanathan}, and C.~{Chae}, ``{SMIET}: Simultaneous molecular
  information and energy transfer,'' \emph{IEEE Wireless Commun.}, vol.~25,
  no.~1, pp. 106--113, Feb. 2018.

\bibitem{arjmandi17}
H.~{Arjmandi}, M.~{Movahednasab}, A.~{Gohari}, M.~{Mirmohseni},
  M.~{Nasiri-Kenari}, and F.~{Fekri}, ``{ISI}-avoiding modulation for
  diffusion-based molecular communication,'' \emph{IEEE Trans. Mol. Biol.
  Multi-Scale Commun.}, vol.~3, no.~1, pp. 48--59, Mar. 2017.

\bibitem{leeson12}
M.~S. Leeson and M.~D. Higgins, ``Forward error correction for molecular
  communications,'' \emph{Nano Commun. Networks}, vol.~3, no.~3, pp. 161--167,
  2012.

\bibitem{shih13}
P.~{Shih}, C.~{Lee}, P.~{Yeh}, and K.~{Chen}, ``Channel codes for reliability
  enhancement in molecular communication,'' \emph{IEEE J. Sel. Areas Commun.},
  vol.~31, no.~12, pp. 857--867, Dec. 2013.

\bibitem{tepekule15}
B.~{Tepekule}, A.~E. {Pusane}, H.~B. {Yilmaz}, C.~{Chae}, and T.~{Tugcu},
  ``{ISI} mitigation techniques in molecular communication,'' \emph{IEEE Trans.
  Mol. Biol. Multi-Scale Commun.}, vol.~1, no.~2, pp. 202--216, Jun. 2015.

\bibitem{noel14}
A.~{Noel}, K.~C. {Cheung}, and R.~{Schober}, ``Improving receiver performance
  of diffusive molecular communication with enzymes,'' \emph{IEEE Trans.
  Nanobiosci.}, vol.~13, no.~1, pp. 31--43, Mar. 2014.

\bibitem{heren15}
A.~C. {Heren}, H.~B. {Yilmaz}, C.~{Chae}, and T.~{Tugcu}, ``Effect of
  degradation in molecular communication: Impairment or enhancement?''
  \emph{IEEE Trans. Mol. Biol. Multi-Scale Commun.}, vol.~1, no.~2, pp.
  217--229, Jun. 2015.

\bibitem{kristensen11}
A.~S. Kristensen, J.~Andersen, T.~N. J{\o}rgensen, L.~S{\o}rensen, J.~Eriksen,
  C.~J. Loland, K.~Str{\o}mgaard, and U.~Gether, ``{SLC6} neurotransmitter
  transporters: Structure, function, and regulation,'' \emph{Pharmacol. Rev.},
  vol.~63, no.~3, pp. 585--640, 2011.

\bibitem{balevi13}
E.~{Balevi} and O.~B. {Akan}, ``A physical channel model for nanoscale
  neuro-spike communications,'' \emph{IEEE Trans. Commun.}, vol.~61, no.~3, pp.
  1178--1187, Mar. 2013.

\bibitem{veletic15}
M.~{Veleti{\'{c}}}, F.~{Mesiti}, P.~A. {Floor}, and I.~{Balasingham},
  ``Communication theory aspects of synaptic transmission,'' in \emph{Proc.
  IEEE Intern. Conf. Commun.}, 2015, pp. 1116--1121.

\bibitem{khan2017}
T.~{Khan}, B.~A. {Bilgin}, and O.~B. {Akan}, ``Diffusion-based model for
  synaptic molecular communication channel,'' \emph{IEEE Trans. Nanobiosci.},
  vol.~16, no.~4, pp. 299--308, Jun. 2017.

\bibitem{sheng07}
M.~Sheng and C.~C. Hoogenraad, ``The postsynaptic architecture of excitatory
  synapses: A more quantitative view,'' \emph{Annu. Rev. Biochem.}, vol.~76,
  no.~1, pp. 823--847, 2007, pMID: 17243894.

\bibitem{zwanzig1991}
R.~Zwanzig and A.~Szabo, ``Time dependent rate of diffusion-influenced ligand
  binding to receptors on cell surfaces,'' \emph{Biophys. J.}, vol.~60, no.~3,
  pp. 671--678, 1991.

\bibitem{berezhkovskii04}
A.~M. Berezhkovskii, Y.~A. Makhnovskii, M.~I. Monine, V.~Y. Zitserman, and
  S.~Y. Shvartsman, ``Boundary homogenization for trapping by patchy
  surfaces,'' \emph{J. Chem. Phys.}, vol. 121, no.~22, pp. 11\,390--11\,394,
  2004.

\bibitem{szabo91}
A.~Szabo, ``Theoretical approaches to reversible diffusion-influenced
  reactions: Monomer--excimer kinetics,'' \emph{J. Chem. Phys.}, vol.~95,
  no.~4, pp. 2481--2490, 1991.

\bibitem{collins1949}
F.~C. Collins and G.~E. Kimball, ``Diffusion-controlled reaction rates,''
  \emph{J. Colloid Sci.}, vol.~4, no.~4, pp. 425--437, 1949.

\bibitem{ahmadzadeh16}
A.~{Ahmadzadeh}, H.~{Arjmandi}, A.~{Burkovski}, and R.~{Schober},
  ``Comprehensive reactive receiver modeling for diffusive molecular
  communication systems: Reversible binding, molecule degradation, and finite
  number of receptors,'' \emph{IEEE Trans. Nanobiosci.}, vol.~15, no.~7, pp.
  713--727, Oct. 2016.

\bibitem{carslaw86}
H.~Carslaw and J.~Jaeger, \emph{Conduction of Heat in Solids}, ser. Oxford
  science publications.\hskip 1em plus 0.5em minus 0.4em\relax Clarendon Press,
  1986.

\bibitem{berezhkovskii13a}
A.~M. Berezhkovskii and A.~Szabo, ``Effect of ligand diffusion on occupancy
  fluctuations of cell-surface receptors,'' \emph{J. Chem. Phys.}, vol. 139,
  no.~12, p. 121910, 2013.

\bibitem{andrews09}
S.~S. Andrews, ``Accurate particle-based simulation of adsorption, desorption
  and partial transmission,'' \emph{Phys. Biol.}, vol.~6, no.~4, p. 046015,
  Nov. 2009.

\bibitem{rice85}
M.~Rice, G.~Gerhardt, P.~Hierl, G.~Nagy, and R.~Adams, ``Diffusion coefficients
  of neurotransmitters and their metabolites in brain extracellular fluid
  space,'' \emph{Neuroscience}, vol.~15, no.~3, pp. 891--902, 1985.

\bibitem{alberts14}
B.~Alberts, D.~Bray, K.~Hopkin, A.~Johnson, J.~Lewis, M.~Raff, K.~Roberts, and
  P.~Walter, \emph{Essential Cell Biology}.\hskip 1em plus 0.5em minus
  0.4em\relax Garland Science, 2014.

\end{thebibliography}
\end{document}